\documentclass{nature}

\usepackage{amssymb,amsmath,graphicx,longtable}
\usepackage{textcomp}
\def\simlt{\mathrel{\hbox{\rlap{\hbox{\lower4pt\hbox{$\sim$}}}\hbox{$<$}}}}
\def\simgt{\mathrel{\hbox{\rlap{\hbox{\lower4pt\hbox{$\sim$}}}\hbox{$>$}}}}

\def\ale{\mathrel{\hbox{\rlap{\hbox{\lower4pt\hbox{$\sim$}}}\hbox{$<$}}}}
\def\age{\mathrel{\hbox{\rlap{\hbox{\lower4pt\hbox{$\sim$}}}\hbox{$>$}}}}

\begin{document}

\title{Reply to ``Rapid $^{14}$C excursion at 3372-3371 BCE not observed at two different locations"}

\author{F. Y. Wang$^{1}$, H. Yu$^2$, Y. C. Zou$^3$, Z. G. Dai$^{1}$ and K. S. Cheng$^4$}

\date{\today}{}
\maketitle

\begin{affiliations}
\item School of Astronomy and Space Science, Nanjing
University, Nanjing 210093, China
\item Department of Astronomy, School of Physics and Astronomy, Shanghai Jiao Tong University, Shanghai, China
\item School of Physics, Huazhong University of Science and Technology, Wuhan 430074, China
\item Department of Physics, University of Hong Kong, Hong Kong, China
\end{affiliations}

\begin{abstract}
The nuclide $^{14}$C can be produced in the atmosphere by high
energy particles and $\gamma$-rays from high-energy phenomena.
Through the carbon cycle, some of $^{14}$CO$_2$ produced in the
atmosphere can be retained in annual tree
rings\cite{Damon95,Damon00}. Four events of rapid increase of the
$^{14}$C content occurred in AD 775, AD 994, BC 660 and BC 3371 were
found\cite{Miyake12,Miyake13,Park17,Wang17}. Recently, the data of
Jull et al. (2020) was inconsistent with our records around BC 3371.
We measured our sample again and found the $^{14}$C records are
consistent with the value in Wang et al. (2017). Therefore, our
$^{14}$C records are robust. The inconsistency may be caused by the
difference of calendar ages for the wood samples, or the physical
origin of the event. First, crossdating on ring width can be
performed only between trees whose growth has the same environmental
conditions. Because the master tree-ring for dendrochronology is
lack for Chinese trees. The master tree-ring from California has to
be used. Therefore, the calendar ages derived from dendrochronology
may be not precise. Second, the $^{14}$C even may be not global. One
evidence is the variation of $^{14}$C content around AD
1006\cite{Damon95,Damon00,Menjo05,Dee17}. The $^{14}$C contents of
Californian trees increase 12\textperthousand~ in two years
\cite{Damon95,Damon00}, while Japanese trees\cite{Menjo05,Dee17}
show no $^{14}$C increase.
\end{abstract}

In order to cross-check the measured result, a new piece of sample
was cut from the buried tree used in our 2017 study\cite{Wang17}. We
separated annual rings carefully. The new-cut sample was measured
using the Accelerator Mass Spectrometry (AMS) method at the
Institute of Accelerator Analysis laboratory (IAA). The new measured
results are shown as red circles in Figure 1 and also listed in
Table 1. This figure also contains the measurement results in Wang
et al. (2017)\cite{Wang17}. Considering the measurement error, we
can see that the new results are consistent with those in Wang et
al. (2017)\cite{Wang17}. Therefore, the $^{14}$C records are robust.
The overall variation shows a rapid increase with a gradual decrease
over several years due to the carbon cycle. The solid line is the
best fit for filled circles using the four-box carbon cycle model
with a net $^{14}$C production of $Q=(7.2\pm1.2)\times 10^7$
atoms/cm$^2$. The parameters of the four-box carbon cycle model are
the same as those used in ref.\cite{Miyake12,Wang17}.

The calendar ages for our wood sample and Jull's samples may not be
at the same time as also suggested by the author. The calendar age
is usually determined by dendrochronology (or tree-ring dating).
Dendrochronology is the scientific method of dating tree rings to
the exact year they were formed, which is widely used. In
dendrochronology, master tree-ring is required. In dendrochronology,
crossdating on ring width can be performed only between trees whose
growth has the same environmental conditions. However, there is no
such master tree-ring for about five thousand years in China. So, we
have to use the master tree-ring from other places\cite{noaa}.
Because, the growth rate (width) of trees is controlled by numerous
variables such as climate and atmospheric conditions. The climate
and atmospheric conditions in central China and California are
dramatically different. Therefore, the age of Chinese wingnut tree
can not be precisely determined using the master tree-ring from
California. The dpLR (Dendrochronology Program Library) may not
suitable for correlation analysis of different tree-ring
records\cite{Jull19}, which may cause the unexpected value of
correlation. So the calendar ages derived from dendrochronology may
contain systematic errors, which is unpredictable. Besides
dendrochronology, we also dated the wood sample using the following
method in our 2017 paper\cite{Wang17}. From $^{14}$C measurements, the calibrated ages of
tree rings are given by IntCal13 curve\cite{Reimer13}. If the age of
the innermost ring is $x$, the ages of outer rings can be determined
by the order of tree rings. The least squares method is used to
constrain the value of $x$ by comparing calibrated ages with ages
from the order of tree rings. Using this method, we found that the
rapid $^{14}$C increase occurs at about BC 3371. However, due to the
large error of calibrated ages (i.e., several decades), the derived
age has the same error. Therefore, the rapid $^{14}$C increase
occurred around BC 3371 with an error of several decades. In order
to derive the exact calendar age of the wood sample from
dendrochronology, the master tree-ring from central China is needed.
We hope it will be built in the future.

The physical origins of $^{14}$C events are still unclear. Although,
some works demanded that solar proton events (SPEs) produce the AD
775 event\cite{Mekhaldi15,Buntgen18}, basing on that this event is
global and the $^{10}$Be and $^{36}$Cl data in ice
cores\cite{Miyake19}. However, Sigl et al. (2015) found that the
time offset between $^{14}$C peak and peaks of $^{10}$Be and
$^{36}$Cl is about 7 years\cite{Sigl15}. Mekhaldi et al. (2015)
adjusted the $^{10}$Be and $^{36}$Cl ice-core records to fit the
tree rings 14C peaks by hands\cite{Mekhaldi15}. The $^{10}$Be is
only 2$\sigma$ above the noise\cite{Neuhauser15}. In the yearly
$^{10}$Be data of the last 400 years, there are more deviations of
similar strength (as claimed by Mekhaldi et al. (2015) for AD 775
and 994), e.g. around 1460, 1605, 1865, and 1890 in North Greenland
Ice Core Project, are all unrelated to strong $^{14}$C
variations\cite{Neuhauser15}. Therefore, the peaks of $^{14}$C and
$^{10}$Be may have different origins, which challenges the SPE
origin. Meanwhile, there are also several problems with the
interpretation of these $^{14}$C events as
SPEs\cite{Neuhauser14,Neuhauser15,Wang19}, including no definite
historic records of strong aurorae or sunspots around AD 775 and AD
994\cite{Stephenson14,Chai15}, the inferred solar fluence ($>$30
MeV) value is inconsistent with the occurrence probability
distribution for SPEs\cite{Cliver14,Usoskin17}, and whether the Sun
can produce such large proton events is still
debated\cite{Schrijver12}. Neuh\"{a}user \& Hambaryan (2014)
calculated the probability for one solar superflare with energy
larger than $10^{35}$ erg within 3000 years to be possibly as low as
0.3 to 0.008. For the BC 3371 event, there is no $^{10}$Be and
$^{36}$Cl measurement at present. There is no reason to claim that
the four events are all caused by SPEs. The Carrington event is the
most energetic solar flare observed so far, which implies that
upsurges of greater magnitude may require extra-solar explanations.
Meanwhile, gamma-ray photons produced by high-energy explosions can
also generate $^{14}$C events, such as
supernovae\cite{Damon95,Miyake12}, short gamma-ray
bursts\cite{Hambaryan13,Pavlov13}, and pulsar
outbursts\cite{Wang19}. A supernova remnant, named Vela Jr., has an
age of $t = 2,000-13,000$ yr at a distance less than 400
parsecs\cite{Allen15}, which is a promising candidates to cause the
BC 3371 event. It has long been hypothesized that intense bursts of
high-energy gamma-ray flux would also be accompanied by ozone
depletion, on account of changing the exchanges in the
atmosphere\cite{Ruderman74,Gehrels03,Pavlov13}. The effect on
atmosphere by gamma-ray radiation is poorly understood and needed
extensive study. This will affect the $^{14}$C absorption rate in
different locations. So the tree rings at the same period may have
different $^{14}$C contents in different regions. One possible
evidence is the $^{14}$C increase around AD 1006. Damon et al.
(1995) measured annual samples of sequoia wood from America between
the years AD 995-1020. They found that the $^{14}$C content
increases from -22.2\textperthousand~ (AD 1009) to
-10.0\textperthousand~ (AD 1011)\cite{Damon95,Damon00}. The $^{14}$C
content increases about 12\textperthousand~ in two years. Meanwhile,
an increase was also found in the IntCal98 curve in the same
period\cite{Stuiver98}. However, Menjo et al. (2005) measured wood
samples from Japan\cite{Menjo05} during the same time-period. No
rapid $^{14}$C increase was found around AD 1006, and the measured
$^{14}$C records are dramatically different for the three wood
samples (see table 2 of Dee et al. 2017)\cite{Dee17}. The possible
reason is the regional effect, which can cause different $^{14}$C
variations in different regions. Although, whether the rapid
increase around AD 1006 is produced by SN 1006 is unclear. The
different $^{14}$C records confirm the regional effect of $^{14}$C
contents during the same period.

In conclusion, the $^{14}$C records in Wang et al. (2017) are
robust. The difference between Wang et al. (2017) and Jull et al.
(2020) may be caused by the difference of calendar ages for the wood
samples, or the increase may be not global. The master tree-ring in
China is required to determine the age of sample precisely.
Meanwhile, study the true origin of $^{14}$C events is important.

{\bf Methods}

{\bf AMS measurement at IAA.} The sample was neutralized with pure
water, and dried. In the acid treatments, the sample is treated with
HCl (1 M). In the standard alkaline treatment, the sample is treated
with NaOH, by gradually raising the concentration level from 0.001 M
to 1 M. If the alkaline concentration reaches 1 M during the
treatment, the treatment is described as Acid-Alkali-Acid (AAA),
while ``AaA" if the concentration does not reach 1 M. (3) The sample
was oxidized by heating to produce CO$_2$ gas. (4) The CO$_2$ gas
was purified in a vacuum line. (5) The purified CO$_2$ gas was
reduced to graphite by hydrogen using iron as a catalyst. (6) The
graphite was pressed into a target holder for the AMS $^{14}$C
dating. Measurement The graphite sample is measured against a
standard of Oxalic acid provided by the National Institute of
Standards and Technology, using a $^{14}$C-AMS system based on the
tandem accelerator. The background check was also performed.

\noindent \large{\bf Author Contributions} All authors discussed and wrote the reply. Corresponding authors
F.Y.W. (fayinwang@nju.edu.cn) and Y.C.Z. (zouyc@hust.edu.cn).

\noindent \large{\bf Competing interests statement} The authors
declare that they have no competing financial interests.

    \begin{table}
    \centering
        \begin{tabular}{|c|c|c|}
 \hline
         Year (BC)   & $\Delta$$^{14}$C(\textperthousand) &  Error$^a$  \\
            \hline
            3375    & 67.02   & 1.90     \\
            \hline
            3373    & 61.06   & 1.90    \\
            \hline
            3372    & 63.97   & 2.00     \\
            \hline
            3371    & 72.60   & 2.00     \\
            \hline
            3370    & 73.23   & 1.90    \\
            \hline
            3369    & 71.96   & 1.80    \\
            \hline
            3368    & 68.22   & 1.90    \\
            \hline
            3366    & 66.12   & 1.90    \\
            \hline
            3364    & 65.23   & 1.90    \\
            \hline
            3361    & 66.99   & 1.80    \\
            \hline
        \end{tabular}\\
        $^a$ The error (s.d.) of $\Delta$$^{14}$C is calculated from error propagation.
        \caption{Measured results in the Institute of Accelerator
Analysis laboratory.}
        \label{tab:tab1}
    \end{table}

\clearpage

\begin{figure}
\centering
\includegraphics[width=0.8\textwidth]{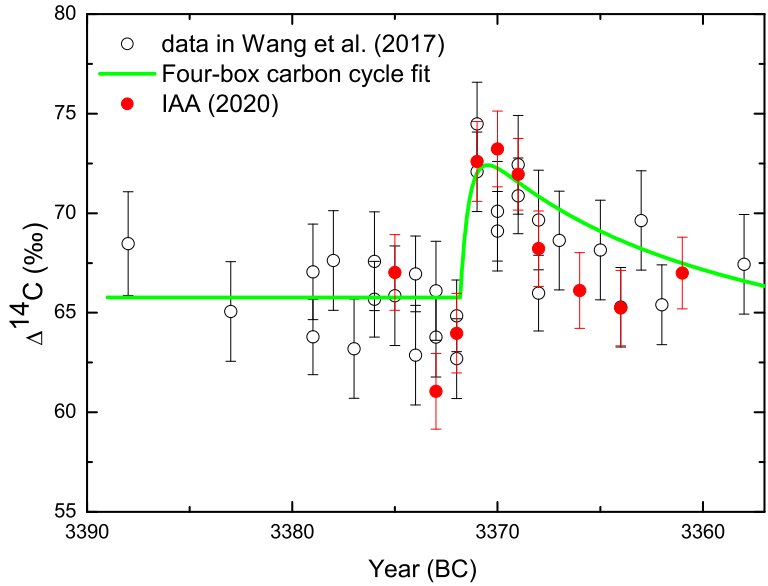}
\caption{\label{Fig1}{\bf Measured $^{14}$C content.} Measured
results of $\Delta$$^{14}$C for the tree rings using the AMS method
at the Institute of Accelerator Analysis laboratory in 2020 (red
circles). The measured results in Wang et al. (2017) are also shown
as black circles. The solid line is the best fit for all data using
the four-box carbon cycle model with a net $^{14}$C production of
$Q=(7.2\pm1.2)\times10^7\rm\,atoms/cm^{2}$. }
\end{figure}

\end{document}